\documentstyle[12pt,epsf]{article}
\catcode`@=11
\def\chkspace{%
  \relax   
  \begingroup\ifhmode\aftergroup\dochksp@ce\fi\endgroup}
\def\dochksp@ce{%
  \unskip              
  \futurelet\chkspct@k\d@chkspc  
}
\def\d@chkspc{%
  \let\nxtsp@ce=\relax
  \ifx\chkspct@k.\else     
    \ifx\chkspct@k,\else
      \ifx\chkspct@k;\else
        \ifx\chkspct@k!\else
          \ifx\chkspct@k?\else
            \ifx\chkspct@k:\else
              \ifx\chkspct@k)\else
              \ifx\chkspct@k(\else
                \ifx\chkspct@k]\else
                  \ifx\chkspct@k-\else
                    \ifx\chkspct@k\egroup\else  
                      \let\nxtsp@ce=\put@space  
                    \fi
                  \fi
                \fi
              \fi
              \fi
            \fi
          \fi
        \fi
      \fi
    \fi
  \fi
  \nxtsp@ce
}
\def\put@space{$\;$}
\catcode`@=12

\def\ra{\relax\ifmmode \rightarrow\else{{$\rightarrow$}}\fi\chkspace}
\def\Ra{\relax\ifmmode \Rightarrow\else{{$\Rightarrow$}}\fi\chkspace}
\def\etal{{\it et al.}\chkspace}

\def\ie{{\it i.e.}\chkspace}

\def\ep{{e$^+$e$^-$}\chkspace}

\def\qu{\relax\ifmmode \quad\else{{$\quad$}}\fi\chkspace}

\def\gluino{\relax\ifmmode \tilde{g} \else $\tilde{g}$ \fi\chkspace}

\def\qq{\relax\ifmmode q\overline{q}
\else $q\overline{q}$ \fi\chkspace}
\def\ff{\relax\ifmmode f\overline{f}
\else $f\overline{f}$ \fi\chkspace}

\def\bb{\relax\ifmmode b\bar{b}
       \else $b\bar{b}$ \fi\chkspace}
\def\cc{\relax\ifmmode { c}\bar{ c}
       \else ${ c}\bar{ c}$ \fi\chkspace}
\def\ccrm{\relax\ifmmode {\rm c}\bar{\rm c}
       \else ${\rm c}\bar{\rm c}$ \fi\chkspace}
\def\tt{\relax\ifmmode {\rm t}\bar{\rm t}
       \else t$\bar{\rm t}$ \fi\chkspace}
\def\ss{\relax\ifmmode {\rm s}\bar{\rm s}
       \else ${\rm s}\bar{\rm s}$ \fi\chkspace}
\def\uu{\relax\ifmmode {\rm u}\bar{\rm u}
       \else ${\rm u}\bar{\rm u}$ \fi\chkspace}
\def\dd{\relax\ifmmode {\rm d}\bar{\rm d}
       \else ${\rm d}\bar{\rm d}$ \fi\chkspace}
\def\QQ{\relax\ifmmode Q\overline{Q}
       \else $Q\overline{Q}$ \fi\chkspace}

\def\qqg{\relax\ifmmode q\overline{q}g
\else $q\overline{q}g$ \fi\chkspace}
\def\bbg{\relax\ifmmode b\overline{b}g
\else $b\overline{b}g$ \fi\chkspace}
\def\ccg{\relax\ifmmode c\overline{c}g
\else $c\overline{c}g$ \fi\chkspace}
\def\ttg{\relax\ifmmode t\overline{t}g
\else $t\overline{t}g$ \fi\chkspace}

\def\afb{\relax\ifmmode A_{FB} \else
{{$A_{FB}$}}\fi\chkspace}
\def\afbb{\relax\ifmmode A_{FB}^b \else
{{$A_{FB}^b$}}\fi\chkspace}
\def\pafb{\relax\ifmmode \tilde{A}_{FB} \else
{{$\tilde{A}_{FB}$}}\fi\chkspace}
\def\pafbb{\relax\ifmmode \tilde{A}_{FB}^b \else
{{$\tilde{A}_{FB}^b$}}\fi\chkspace}

\def\pafbzo{\relax\ifmmode \tilde{A}_{FB}|_{O(0)} \else
{{$\tilde{A}_{FB}|_{O(0)}$}}\fi\chkspace}
\def\pafbfo{\relax\ifmmode \tilde{A}_{FB}|_{\oalp} \else
{{$\tilde{A}_{FB}|_{\oalp}$}}\fi\chkspace}
\def\pafbso{\relax\ifmmode \tilde{A}_{FB}|_{\oalpsq} \else
{{$\tilde{A}_{FB}|_{\oalpsq}$}}\fi\chkspace}
\def\pafbto{\relax\ifmmode \tilde{A}_{FB}|_{\oalpc} \else
{{$\tilde{A}_{FB}|_{\oalpc}$}}\fi\chkspace}

\def\pafbbzo{\relax\ifmmode \tilde{A}_{FB}^b|_{O(0)} \else
{{$\tilde{A}_{FB}^b|_{O(0)}$}}\fi\chkspace}
\def\pafbbfo{\relax\ifmmode \tilde{A}_{FB}^b|_{\oalp} \else
{{$\tilde{A}_{FB}^b|_{\oalp}$}}\fi\chkspace}
\def\pafbbso{\relax\ifmmode \tilde{A}_{FB}^b|_{\oalpsq} \else
{{$\tilde{A}_{FB}^b|_{\oalpsq}$}}\fi\chkspace}
\def\pafbbto{\relax\ifmmode \tilde{A}_{FB}^b|_{\oalpc} \else
{{$\tilde{A}_{FB}^b|_{\oalpc}$}}\fi\chkspace}

\def\afbo0{\tilde{A}_{FB}|_{O(0)}}
\def\afbo1{\tilde{A}_{FB}|_{\oalp}}
\def\afbo2{\tilde{A}_{FB}|_{\oalpsq}}
\def\afbo3{\tilde{A}_{FB}|_{\oalpc}}

\def\lam{\relax\ifmmode \Lambda_{\overline{MS}}
       \else {{$\Lambda_{\overline{MS}}$}}\fi\chkspace}
\def\lamuds{\relax\ifmmode \Lambda^{(3)}_{\overline{MS}}
       \else {{$\Lambda^{(3)}_{\overline{MS}}$}}\fi\chkspace}
\def\lamudsc{\relax\ifmmode \Lambda^{(4)}_{\overline{MS}}
       \else $\Lambda^{(4)}_{\overline{MS}}$\fi\chkspace}
\def\lamudscb{\relax\ifmmode \Lambda^{(5)}_{\overline{MS}}
       \else $\Lambda^{(5)}_{\overline{MS}}$\fi\chkspace}

\def\alp{\relax\ifmmode \alpha_s\else $\alpha_s$\fi\chkspace}
\def\alpbar{\relax\ifmmode \bar{\alpha_s}
       \else $\bar{\alpha_s}$\fi\chkspace}
\def\alpmz{\relax\ifmmode \alpha_s(M_Z)\else $\alpha_s(M_Z)$\fi\chkspace}
\def\alpmzsq{\relax\ifmmode \alpha_s(M_Z^2)
       \else $\alpha_s(M_Z^2)$\fi\chkspace}

\def\oalp{\relax\ifmmode O(\alpha_s)\else{{O($\alpha_s$)}}\fi\chkspace}
\def\oalpsq{\relax\ifmmode O(\alpha_s^2)
           \else{{O($\alpha_s^2$)}}\fi\chkspace}
\def\oalpc{\relax\ifmmode O(\alpha_s^3)
           \else{{O($\alpha_s^3$)}}\fi\chkspace}
\def\oalpf{\relax\ifmmode O(\alpha_s^4)
           \else{{O($\alpha_s^4$)}}\fi\chkspace}

\def\rb{\relax\ifmmode R_3^b/R_3^{all}
           \else{{$R_3^b/R_3^{all}$}}\fi\chkspace}
\def\rc{\relax\ifmmode R_3^c/R_3^{all}
           \else{{$R_3^c/R_3^{all}$}}\fi\chkspace}
\def\ruds{\relax\ifmmode R_3^{uds}/R_3^{all}
           \else{{$R_3^{uds}/R_3^{all}$}}\fi\chkspace}
\def\ri{\relax\ifmmode R_3^i/R_3^{all}
           \else{{$R_3^i/R_3^{all}$}}\fi\chkspace}
\def\rj{\relax\ifmmode R_3^j/R_3^{all}
           \else{{$R_3^j/R_3^{all}$}}\fi\chkspace}
\def\alpi{\relax\ifmmode \alpha^i_s/\alpha^{all}_s
           \else{{$\alpha^i_s/\alpha^{all}_s$}}\fi\chkspace}

\def\mbz{\relax\ifmmode m_b(M_Z)
           \else{{$m_b(M_Z)$}}\fi\chkspace}
\def\mbb{\relax\ifmmode m_b(M_b)
           \else{{$m_b(M_b)$}}\fi\chkspace}

\def\prl{Phys. Rev. Lett.\chkspace}

\def\z0{\relax\ifmmode Z^0 \else {$Z^0$} \fi\chkspace}
\def\h0{\relax\ifmmode H^0 \else {$H^0$} \fi\chkspace}
\def\Dst{\relax\ifmmode {\rm D}^* \else {D$^*$}\fi\chkspace}
\def\Dpl{\relax\ifmmode {\rm D}^+ \else {D$^+$}\fi\chkspace}
\def\D0{\relax\ifmmode {\rm D}^0 \else {D$^0$}\fi\chkspace}
\def\Kst{\relax\ifmmode {\rm K}^* \else {K$^*$}\fi\chkspace}
\def\K0{\relax\ifmmode {\rm K}^0_s \else {K$^0_s$}\fi\chkspace}
\def\Kpl{\relax\ifmmode {\rm K}^+ \else {K$^+$}\fi\chkspace}
\def\Kstz{\relax\ifmmode {\rm K}^{*0} \else {K$^{*0}$}\fi\chkspace}

\def\beq{\begin{equation}}
\def\eeq{\end{equation}}
\def\bea{\begin{eqnarray}}
\def\eea{\end{eqnarray}}

\renewcommand{\baselinestretch}{1.5}

 1

\catcode`\@=11 
\makeatletter
\def\@seccntformat#1{\csname the#1\endcsname.\hskip 1em}

\makeatother
\begin{document}

\thispagestyle{empty}
\begin{flushright}
{\footnotesize\renewcommand{\baselinestretch}{.75}
SLAC--PUB--8737\\
February 2001\\
}
\end{flushright}

\vskip 1truecm
 
\begin{center}
{\Large \bf 
IMPROVED MEASUREMENT OF THE\\
PROBABILITY FOR GLUON SPLITTING\\ 
INTO \bb IN \z0 DECAYS$^*$}
 
\vspace {1.0cm}
 
 {\bf The SLD Collaboration$^{**}$}\\
Stanford Linear Accelerator Center \\
Stanford University, Stanford, CA~94309
 
\end{center}

\vskip 2truecm

\normalsize
 
\begin{center}
{\bf ABSTRACT }
\end{center}

\small
 
\noindent
We have measured gluon splitting into bottom quarks, 
$g$ \ra \bb, in hadronic $Z^0$ decays collected by SLD between 1996 and 1998.
The analysis was performed by looking for secondary bottom production in
4-jet events of any primary flavor.  
4-jet events were identified, and in each event 
a topological vertex-mass technique was applied to 
the two jets closest in angle in order
to identify them as $b$ or $\overline{b}$ jets. 
The upgraded CCD-based vertex detector gives very high $B$-tagging
efficiency, especially for $B$ hadrons with the low energies typical of
this process.
We measured the rate of $g$ \ra \bb production
 per hadronic event, $g_{b\bar{b}}$, to be 
$(2.44\pm0.59{\rm (stat.)}\pm0.34{\rm (syst.)})\times10^{-3}$.
 
\vskip 1truecm

\centerline{({\it Submitted to Physics Letters B})}

\vfill
{\footnotesize
\noindent
$*$ Work supported by Department of Energy contract DE-AC03-76SF00515 (SLAC).}

\eject

\section{Introduction}  

\vskip .5truecm

\noindent
The vertex representing a gluon splitting into a heavy-quark pair, $g$ 
\ra \QQ~($Q$ = $b$ or $c$), 
is a fundamental elementary component of Quantum Chromodynamics (QCD), 
but the contribution of this vertex to physical processes is 
poorly known, both theoretically and experimentally.
In high-energy \ep annihilation the leading-order process containing
this vertex is \ep \ra \qqg \ra \qq\QQ~($q$ = $u,d,s,c$ or $b$).
Information on $g$ \ra \QQ can
hence be obtained by studying \ep \ra hadrons events comprising four
quark~\cite{quark} jets, with two of the jets identified as $Q$ or $\overline{Q}$.
Background events of the kind \ep \ra \QQ$g$ \ra \QQ\qq are in principle
indistinguishable final states, although their kinematics are typically
quite different: the $Q$ and $\overline{Q}$ jets in this process 
tend to be back-to-back, and 
have energy comparable with the beam energy. By contrast, 
the $Q$ and $\overline{Q}$ jets from $g$ \ra \QQ tend to be collinear, and have
low energy. In order to enhance the contribution from the latter process
it is necessary to identify four-jet events and to tag typically the two jets 
closest in angle, and/or of lowest energy,
as $Q$ or $\overline{Q}$.

We define the rate $g_{Q\bar{Q}}$ as the fraction of \ep \ra hadrons
events in which a gluon splits into \QQ,
\ep \ra \qqg \ra \qq\QQ.
Since the quark mass provides a natural cutoff 
$g_{Q\bar{Q}}$ is an infrared finite quantity, 
which  can be computed in the framework of perturbative QCD.
At the \z0 resonance energy the production of secondary \cc or \bb via
gluon splitting is strongly suppressed by the required large gluon
virtual mass.
From a leading-order + 
next-to-leading-logarithm approximation calculation one expects~\cite{seymour} 
$g_{c\bar{c}}$ to be at the per cent level, and
$g_{b\bar{b}}$ to be at the 0.1 per cent level.
The calculations, however, depend on \alp and on the
quark mass, which results in substantial theoretical uncertainties.

Here we consider the process $g$ \ra \bb.
The measurement of $g_{b\bar{b}}$ is difficult experimentally
since the rate is intrinsically low and 
the backgrounds from \z0 \ra \bb events are two orders of
magnitude larger. In addition, 
the $B$ hadrons from $g$ \ra \bb have relatively low energy and  
short flight distance and are difficult to identify using 
standard tagging techniques.
So far, measurements of $g_{b\bar{b}}$
have been reported by ALEPH, DELPHI and OPAL~\cite{GBBLEP}; the
most precise measurement, from ALEPH, is  $g_{b\bar{b}}$ =
$(2.77\pm0.71)\times10^{-3}$.

Such limited knowledge of $g_{b\bar{b}}$ results in the
main source of uncertainty in the measurement of the partial decay
width  
$R_{b}=\Gamma (Z^0 \to \bb)/ \Gamma (Z^0 \to \qq)$ \cite{Rb},
which is potentially sensitive, via loop effects, to new physics processes that couple
to the $b$-quark. Hence, more precise measurements of $g_{b\bar{b}}$ would 
help to improve the precision of tests of the electroweak theory 
in the heavy-quark  
sector. In addition, knowledge of the $g$ \ra \bb process is vital for measurements at
hadron colliders. For example, about $50$\% 
of the $B$ hadrons produced in QCD processes at the Tevatron are due to
$g$ \ra \bb, and a larger fraction is expected to contribute at the LHC.
These events form a large background to possible rare new processes involving 
decays to heavy quarks, such as $H^0$ \ra \bb, and improved
understanding of $g$ \ra \bb will help to constrain 
background heavy-flavor production at hadron colliders.

We present a measurement of $g_{b\bar{b}}$ based on the sample of 
roughly 400,000 hadronic \z0 decays produced in \ep annihilations 
at the Stanford Linear Collider (SLC) between 1996 and 1998
and collected in the SLC Large Detector (SLD).
In this period, $Z^0$ decays were collected 
with an upgraded vertex detector with
wide acceptance and excellent impact parameter resolution, 
thus improving considerably our tagging capability for the low-energy
$B$ hadrons characteristic of the $g$ \ra \bb process.

\vskip 1truecm

\section{The SLD}
 
\vskip .5truecm

\noindent
A description of the SLD is given elsewhere \cite{SLD}.
Only the details most relevant to this analysis are mentioned here.
The trigger and selection criteria
for \z0 \ra hadrons events are described elsewhere~\cite{sld96}.
This analysis used charged tracks measured in the Central Drift
Chamber (CDC)~\cite{cdc}
and in the upgraded CCD Vertex Detector (VXD)~\cite{vxd3}, with a
momentum resolution of
$\sigma_{p_{\perp}}/p_{\perp}$ = $0.01\oplus0.0026p_{\perp}$,
where $p_{\perp}$ is the track transverse momentum with respect
to the beamline, in GeV/$c$. 
For high-momentum tracks the measured impact-parameter resolution approaches
$7.7\mu$m ($9.6\mu$m) in the plane transverse to (containing) the beamline,
while multiple scattering 
contributions are $29\mu{\rm m}/(p\sin^{3/2}\theta)$ in
both projections.
In \ep \ra hadrons events the centroid of the SLC interaction point (IP) 
was reconstructed with a precision of 
approximately 5$\mu$m (10$\mu$m). 

Only well-reconstructed
tracks~\cite{sld2} were used for $B$-hadron tagging, and
tracks from identified $\gamma$ conversions and $K^0$ or $\Lambda^0$ decays
were removed from consideration. Each track was required to have:
a polar angle satisfying $|\cos \theta|$ $<$ 0.87,
a transverse impact parameter $<$ 0.30cm 
with an error $<$ 250$\mu$m,
impact parameters, measured in the CDC only, of $<$ 1.0cm (transverse)
and $<$ 1.5cm (plane containing the beamline),
at least 23 hits in the CDC
with the first hit $<$ 50.0cm from the IP,
a $\chi^2/d.o.f.$  $<$ 8.0 for the CDC-only and CDC+VXD fits,
and $p_{\perp}$ $>$ 0.25GeV/c.

For the purpose of estimating the efficiency and purity of 
the $g$ \ra \bb selection procedure, 
we made use of a detailed Monte-Carlo simulation of the detector.
The JETSET 7.4 \cite{Sjostrand:1994yb} event generator was used, 
with parameter values tuned to hadronic $e^+e^-$ annihilation 
data \cite{LUNDTUNE}, 
combined with a simulation of $B$ hadron decays  
tuned to $\Upsilon(4S)$ data \cite{SLDSIM} 
and a simulation of the SLD based on GEANT 3.21 \cite{GEANT}.
Inclusive distributions of single-particle and event-topology observables
in hadronic events were found to be well described by the
simulations \cite{SLDALPHAS}.
Uncertainties in the simulation
were taken into account in the systematic errors (Section~\ref{SEC:SYS}).

Monte-Carlo events were reweighted to take into account the current 
measurements of gluon splitting into heavy-quark pairs \cite{GBBLEP,GCCLEP}.
JETSET with the SLD parameters 
predicts $g_{b\bar{b}}=0.14\%$ and $g_{c\bar{c}}=1.36\%$.
We reweighted these events in the simulated sample to obtain 
$g_{b\bar{b}}=0.247\%$ and $g_{c\bar{c}}=3.07\%$~\cite{EPSgbb}.
Samples of about 1900k Monte-Carlo \z0 \ra \qq events,
1900k \z0 \ra \bb events, 1090k \z0 \ra \cc events 
and 60k \z0 \ra \qqg, $g$ \ra \bb events were used
to evaluate the selection efficiencies (Section~\ref{SEC:SYS}).

\vskip 1truecm

\section{Flavor Tagging \label{SEC:BTAG}} 
 
\vskip .5truecm

We used topologically-reconstructed secondary vertices \cite{Jackson:1997sy}
for heavy-quark tagging.
To reconstruct the secondary vertices, the space points 
where track density functions overlap were found in three dimensions.
Only the vertices that are significantly displaced from the IP
were considered to be possible $B$- or $D$-hadron
decay vertices.
The mass of the secondary vertex
was calculated using the tracks that
were associated with the vertex. 
We corrected the reconstructed mass to account for 
neutral decay products and tracks missed from the vertex.
By using kinematic information from
the vertex flight path and
the momentum sum of the tracks associated with the secondary vertex,
we calculated the $P_T$-corrected mass, $M_{P_T}$, 
by adding a component of missing momentum to the invariant mass,
as follows:
$$
M_{P_T} = \sqrt{{M_{vtx}}^2 + P_T^2} + |P_T|.
$$
where $M_{vtx}$ is the invariant mass of the tracks associated with 
the reconstructed secondary vertex and $P_T$ is the total transverse momentum
of the vertex-associated tracks with respect to the vertex axis,
which we estimated
independently of the track momenta by the vector along the
line joining the IP to the reconstructed vertex position.
In this correction, vertexing resolution as well as the IP resolution
are crucial. 

With these features, topological
vertex finding gives excellent $b$-tagging efficiency and purity.
In particular, the efficiency is good even at low $B$-hadron energies,
which is especially important for detecting $g$ \ra \bb. For 
the selected 4-jet event sample (Section~4) we used our
simulation to estimate that our mean tagging efficiency for $b$-jets in
the $g$ \ra \bb process is 67\%.

\vskip 1truecm

\section{Analysis}

\vskip .5truecm

\noindent
Besides the signal events which contain $g$ \ra \bb, hereafter called `B events', 
background events can be divided into two categories:
1) events in which a gluon splits to a charm quark pair, called `C events', and
2) events which do not contain any gluon splitting into heavy quarks 
        at all, hereafter called `Q events'.
In  $g$ \ra \bb events 
the two $B$ hadrons from the gluon splitting tend to be produced with low energy
in a collinear configuration, 
which allows one to discriminate the signal from background.
We first required that each event contain
4 jets, and that the two jets closest in angle were tagged as 
$b$ jets on the basis that each contained a secondary vertex. 
We then examined additional kinematic quantities and used a neural
network technique to improve the signal/background ratio.

In each event jets were formed by applying the Durham jet-finding 
algorithm \cite{Catani:1991hj} with $y_{cut}=0.005$ to the set of 
charged tracks; this $y_{cut}$ value was 
chosen to minimize the sum of the statistical and systematic errors on $g_{b\bar{b}}$.
Events containing four or more jets were retained. 
The jet-finder was re-run on the $>4-$ jet events 
with successively larger $y_{cut}$ values
until exactly four jets were reconstructed.
With this definition the 4-jet rate in the data was $(14.58\pm0.07)\%$, 
where the error is statistical only.
In the Monte-Carlo simulation the 4-jet rate was $(14.47\pm0.02\pm0.16)\%$ 
where the first error is statistical and the second is due to the 
uncertainty in the simulation of heavy-quark physics (Section~\ref{SEC:SYS}).
For the B, C and Q events the 4-jet rates predicted by the simulation are
about $60\%$, $38\%$ and $14\%$, respectively.
Each jet energy was calculated using its associated-track momenta
and assuming all tracks to have the charged pion mass.

In each selected event the two jets closest in angle were considered as
candidates for originating from the gluon splitting process $g$ \ra \bb, and
the topological vertex method was applied to them.
We required both jets to contain a secondary vertex 
with a 3D decay length greater than $300\mu$m.
No tag was applied to the other two jets.
1514 events were selected.
In each event the tagged jets were labeled `1' and `2', where jet 1 contained
the vertex with the
greater $M_{P_T}$ value, $M_{P_T 1}$, and jet 2 that with the lesser
$M_{P_T}$ value, $M_{P_T 2}$.
The other two jets in the event were labeled `3' and `4',
where jet 3 was more energetic than jet 4.
With these requirements the selection efficiency for $g$ \ra \bb events
was estimated to be $16.2\%$,
with a signal/background ratio in the selected sample of
approximately 1/10.
$75\%$ of the background came from \z0 \ra \bb events, 
$9\%$ from \z0 \ra \qq~($q \ne b$) events,
and the remaining $16\%$ from $g$ \ra \cc events. 
In order to improve the signal/background ratio 
we used a neural network technique.
We chose the following 9 observables as inputs to the neural
network; each observable was scaled to correspond to a range 
between 0 and 1.

\begin{figure}[h]       
\centerline{\epsfxsize 2.5 truein \epsfbox{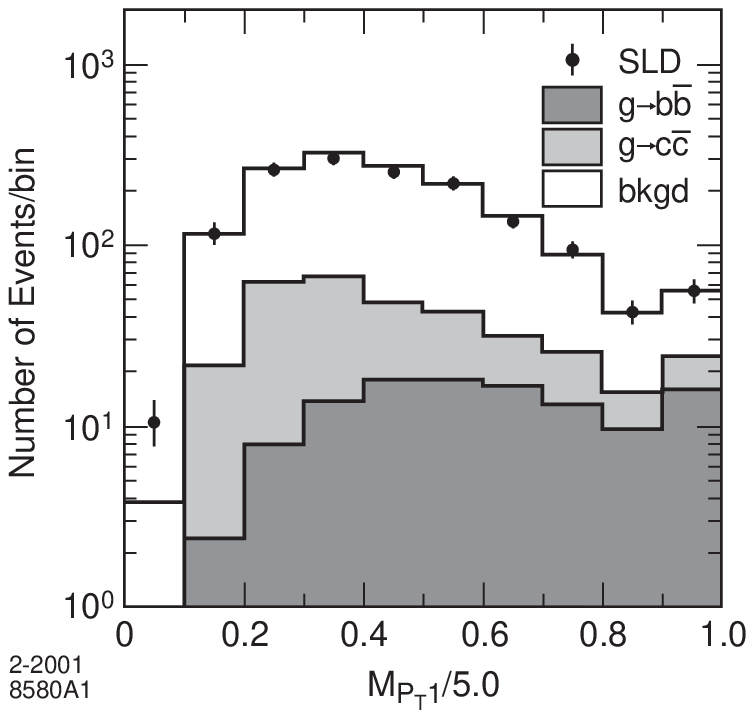}}   
\caption[]{
 Distribution of $M_{P_T 1}$ (see text).
The right-most bin includes overflows.
        The points represent the data, the histogram the simulation; 
 the expected contribution from $g$ \ra \bb~ ($g$ \ra \cc) (see text) 
is shown as the dark (shaded) area.
}
\label{Fig:VMASSMAX}
\end{figure}

\begin{figure}[h]     
\centerline{\epsfxsize 2.5 truein \epsfbox{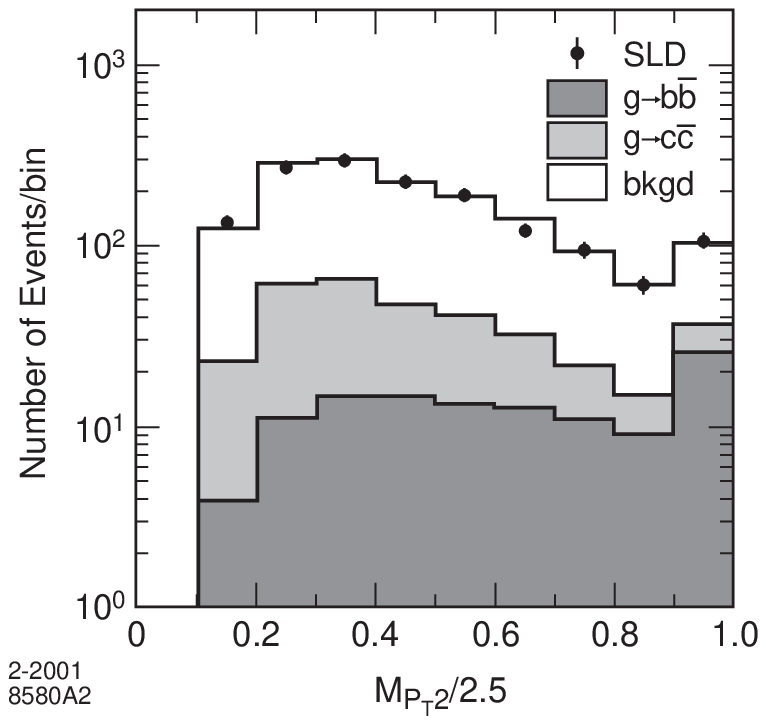}}   
\caption[]{  
Distribution of $M_{P_T 2}$ (see text).
The right-most bin includes overflows.
        The points represent the data, the histogram the simulation; 
 the expected contribution from $g$ \ra \bb~ ($g$ \ra \cc) (see text) 
is shown as the dark (shaded) area.
}
\label{Fig:VMASSMIN}
\end{figure}

\begin{figure}[h]       
\centerline{\epsfxsize 2.5 truein \epsfbox{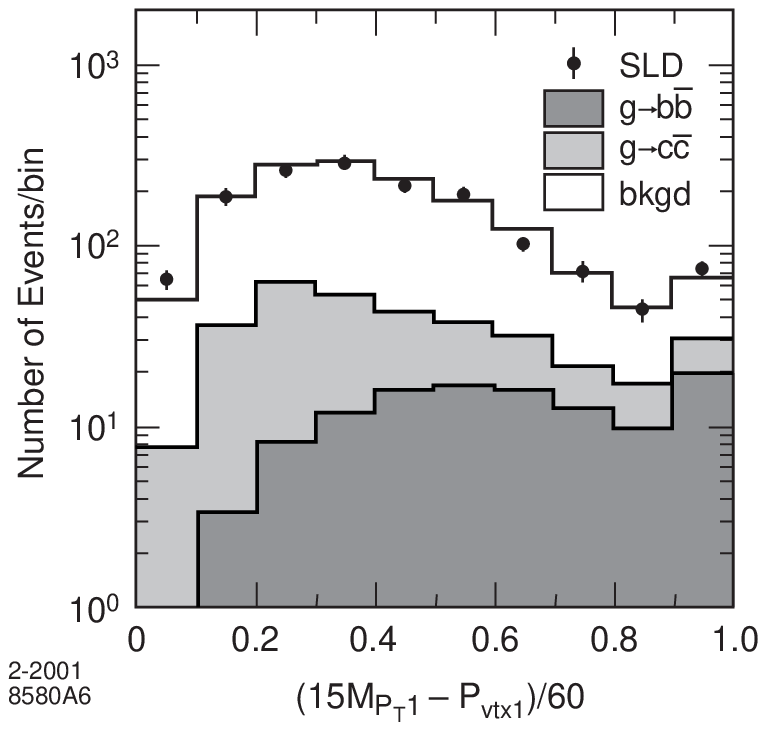}}   
\caption[]{
  Distribution of $15 M_{P_T 1} - P_{vtx 1}$ (see text).
The right-most bin includes overflows.
        The points represent the data, the histogram the simulation; 
 the expected contribution from $g$ \ra \bb~ ($g$ \ra \cc) (see text) 
is shown as the dark (shaded) area.
}
\label{Fig:PVTX1}
\end{figure}

\begin{figure}[h]       
\centerline{\epsfxsize 2.5 truein \epsfbox{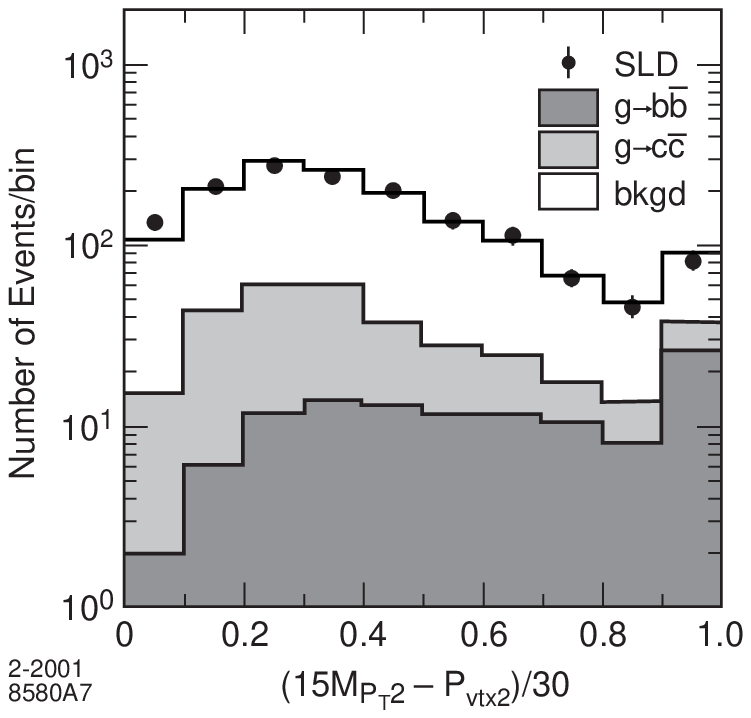}}   
\caption[]{
 Distribution of $15 M_{P_T 2} - P_{vtx 2}$ (see text).
The right-most bin includes overflows.
        The points represent the data, the histogram the simulation; 
 the expected contribution from $g$ \ra \bb~ ($g$ \ra \cc) (see text) 
is shown as the dark (shaded) area.
}
\label{Fig:PVTX2}
\end{figure}

\begin{figure}[h] 
\centerline{\epsfxsize 2.5 truein \epsfbox{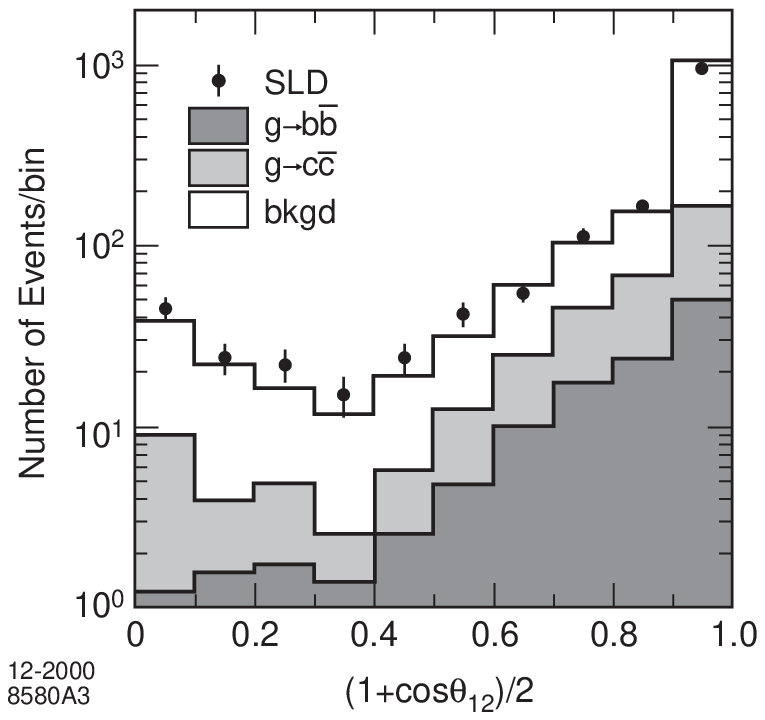}}
\caption[]{
  Distribution of cos$\theta_{12}$ (see text).
        The points represent the data, the histogram the simulation; 
 the expected contribution from $g$ \ra \bb~ ($g$ \ra \cc) (see text) 
is shown as the dark (shaded) area.
}
\label{Fig:COS12}
\end{figure}

\begin{figure}[h]       
\centerline{\epsfxsize 2.5 truein \epsfbox{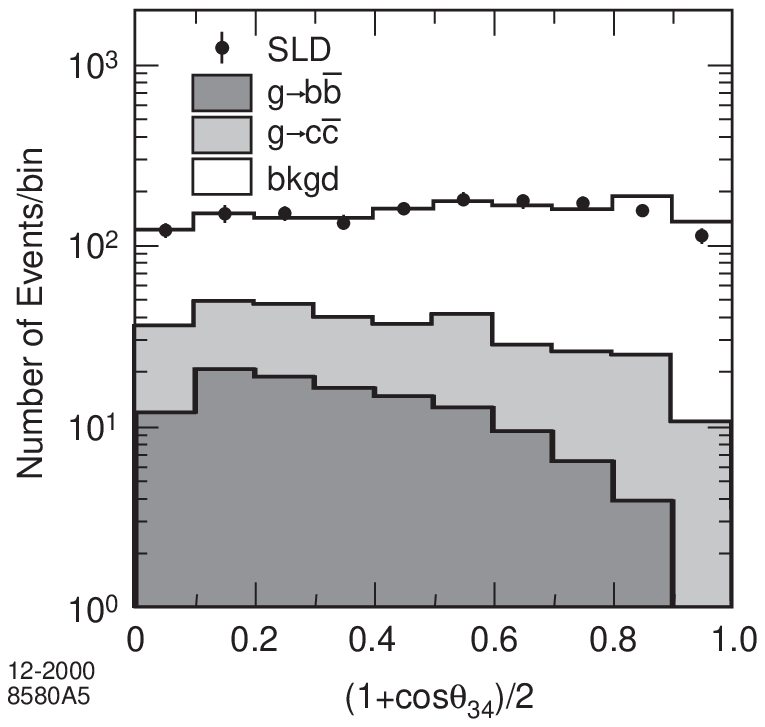}}
\caption[]{
\small  Distribution of cos$\theta_{34}$ (see text).
        The points represent the data, the histogram the simulation; 
 the expected contribution from $g$ \ra \bb~ ($g$ \ra \cc) (see text) 
is shown as the dark (shaded) area.
}
\label{Fig:COS34}
\end{figure}

\begin{figure}[h]       
\centerline{\epsfxsize 2.5 truein \epsfbox{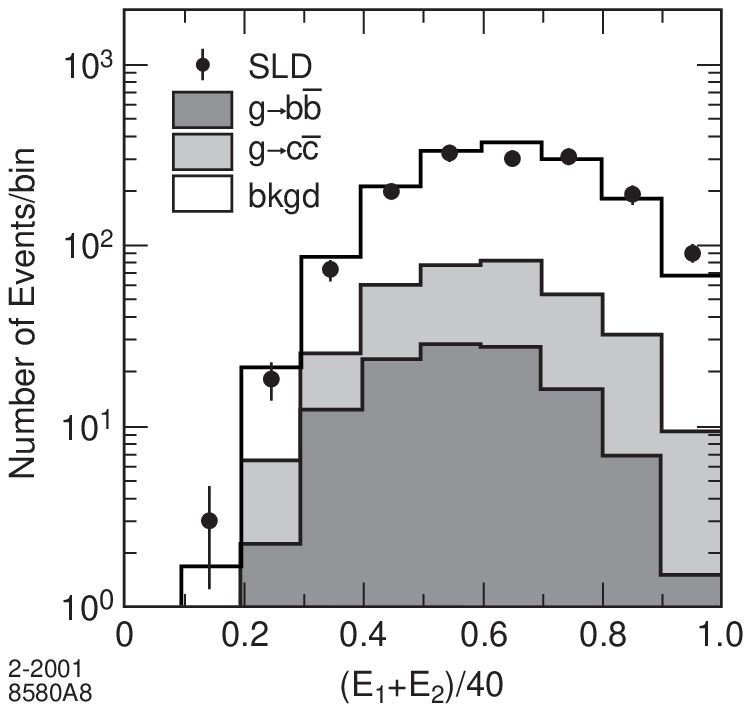}}
\caption[]{
 The distribution of the energy sum of jets 1 and 2.
        The points represent the data, the histogram the simulation; 
 the expected contribution from $g$ \ra \bb~ ($g$ \ra \cc) (see text) 
is shown as the dark (shaded) area.
}                                                
\label{Fig:EJET12}
\end{figure}

\begin{figure}[h]       
\centerline{\epsfxsize 2.5 truein \epsfbox{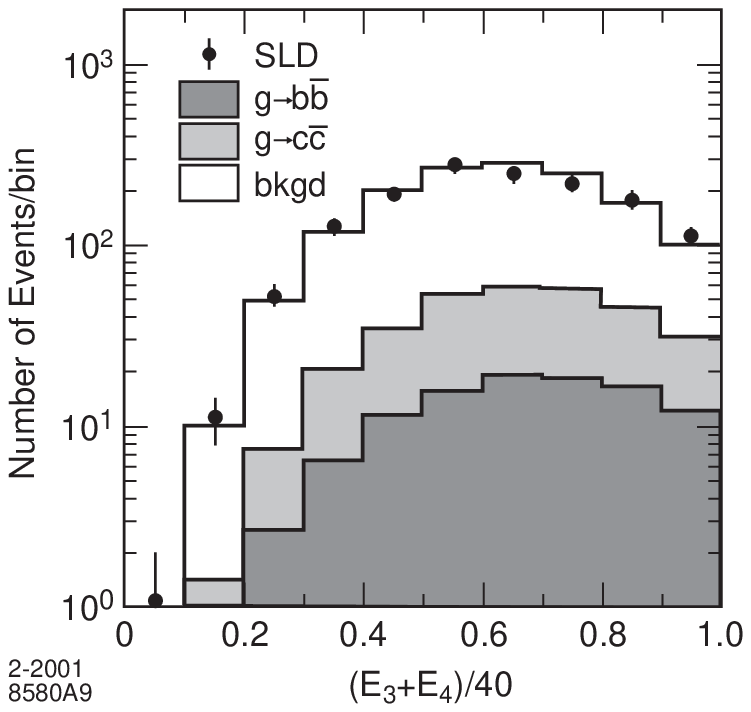}}
\caption[]{
  The distribution of the energy sum of jets 3 and 4.
        The points represent the data, the histogram the simulation; 
 the expected contribution from $g$ \ra \bb~ ($g$ \ra \cc) (see text) 
is shown as the dark (shaded) area.
}
\label{Fig:EJET34}
\end{figure}

\begin{figure}[h]       
\centerline{\epsfxsize 2.5 truein \epsfbox{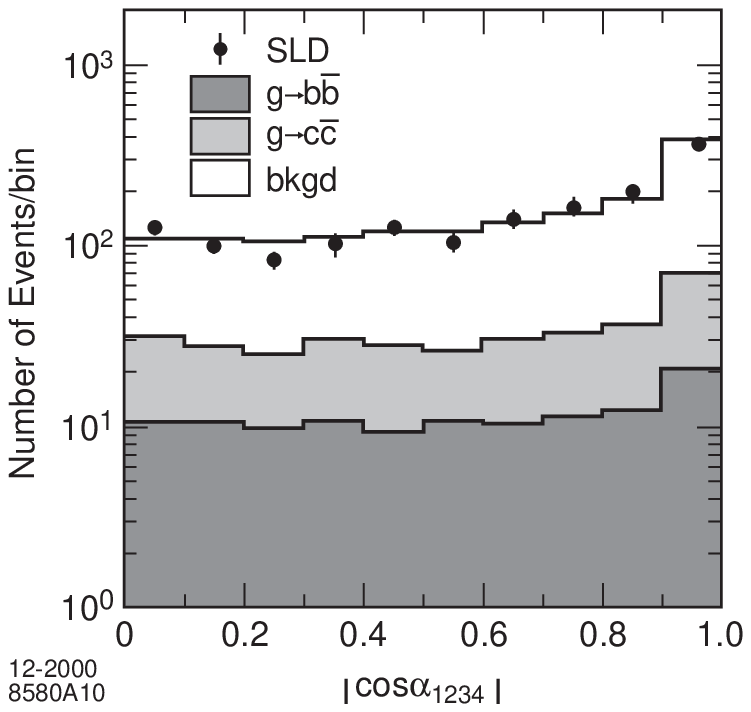}}
\caption[]{
The distribution of $\cos\alpha_{1234}$ (see text).
        The points represent the data, the histogram the simulation; 
 the expected contribution from $g$ \ra \bb~ ($g$ \ra \cc) (see text) 
is shown as the dark (shaded) area.
}
\label{Fig:COS1234}
\end{figure}

\begin{enumerate}

 \item $M_{P_T 1}$:
       $b$ jets typically have higher values of this quantity than
 	$c$ or $uds$ jets. The distribution of $M_{P_T 1}$ is shown in
       Figure~\ref{Fig:VMASSMAX}.

 \item $M_{P_T 2}$: This observable has similar discriminating power.
       The distribution of $M_{P_T 2}$ is shown in
       Figure~\ref{Fig:VMASSMIN}. 

 \item $15 M_{P_T 1} - P_{vtx 1}$: 
	where $P_{vtx}$ is the vertex momentum.
        This observable 
	tends to be large for $b$ jets since $B$ decay vertices typically
        have higher mass than those from charm decays, and
        vertices resulting from $B$ \ra $D$ cascade 
        decays have a lower momentum than
	those from primary $D$ hadrons.
        The distribution of $15 M_{P_T 1} - P_{vtx 1}$ is shown in Figure~\ref{Fig:PVTX1}.
 
 \item $15 M_{P_T 2} - P_{vtx 2}$: 
	This observable also has discriminating power between signal and
	background events.
        The distribution of $15 M_{P_T 2} - P_{vtx 2}$ is shown in 
        Figure~\ref{Fig:PVTX2}. 

 \item  The angle $\theta_{12}$ between  the vertex axes of jets 1 and 2.
        The two jets from $g$ \ra \bb tend to
       have $\theta_{12}$ $\simeq 0$. However, in some
       \z0 \ra \bb background events a single $b$ jet may be split into two
       by the jet-finder; in these cases the two reconstructed vertices
       tend to have $\cos\theta_{12}\geq0.98$.
       The distribution of cos$\theta_{12}$ is shown in Figure~\ref{Fig:COS12}.

 \item The angle $\theta_{34}$ between the axes of jets 3 and 4.
       In events containing $g$ \ra \bb this tends
       to be near $\pi$, while background events tend to
       populate the smaller-angle region.
The distribution of cos$\theta_{34}$ is shown in Figure~\ref{Fig:COS34}.
       
 \item The energy sum of jets 1 and 2, $E_1+E_2$:
       The two jets arising from $g$ \ra \bb tend to have lower energy
       than the other two jets in the event.
The distribution of $E_1+E_2$ is shown in Figure~\ref{Fig:EJET12}.

 \item The energy sum of jets 3 and 4, $E_3+E_4$:
This tends to be larger in signal events than in background events.
The distribution of $E_3+E_4$ is shown in Figure~\ref{Fig:EJET34}.

 \item The angle $\alpha_{1234}$ between 
       the plane $\Pi_{12}$ formed by jets 1 and 2 and the plane $\Pi_{34}$
       formed by jets 3 and 4:
       This variable is similar to the Bengtsson-Zerwas
       angle~\cite{Bengtsson:1988qg}, and is useful to separate 
       $g$ \ra \bb events because the radiated virtual gluon in the
       process \z0 \ra \qqg is polarized in the three-parton plane, 
       and this is reflected in its subsequent
       splitting, by favoring $g$ \ra \qq emission out of
       this plane, \ie $\alpha_{1234}\simeq\pi/2$. 
The distribution of cos$\alpha_{1234}$~\cite{alpha} 
is shown in Figure~\ref{Fig:COS1234}. 

\end{enumerate}

\begin{figure}[h]       
\centerline{\epsfxsize 2.5 truein \epsfbox{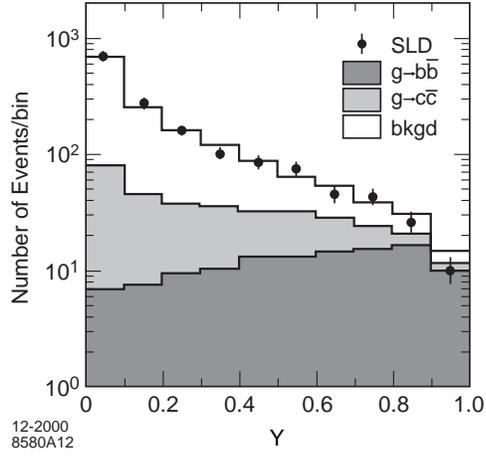}}
\caption[]{
Distribution of the neural network output $Y$.
        The points represent the data, the histogram the simulation; 
 the expected contribution from $g$ \ra \bb~ ($g$ \ra \cc) (see text) 
is shown as the dark (shaded) area.
}
\label{Fig:NNoutput}
\end{figure}

The measured and simulated distributions agree well for these input observables.
We trained the neural network using Monte-Carlo samples of 
about 1800k \z0 \ra \qq events,
1200k \z0 \ra \bb events, 780k \z0 \ra \cc events 
and 50k events containing $g$ \ra \bb.
These samples were independent of the ones used for the
selection efficiency and background studies.
Figure~\ref{Fig:NNoutput} shows the distribution of the neural
network output variable, $Y$.
We retained events with $Y$ greater than 0.7.
This value was found to minimise the total error on the final
$g_{b\bar{b}}$ result. 

\vskip 1truecm

\section{Result}

\noindent
79 events were selected in the data.
The number of background events was estimated, 
using the Monte Carlo simulation, 
to be $41.9$, 
where $35\%$ of the background comes from $g$ \ra \cc events,
$63\%$ from \z0 \ra \bb events, 
and the remaining $2\%$ from \z0 \ra \qq $(q \ne b)$ events.
Table \ref{Table:SELEFF} shows the selection efficiencies,
relative to the selected hadronic-event sample, for the B, C and Q 
event categories.
From these efficiencies and the fraction of events selected in the data,
$f_{d}=(2.73\pm0.31)\times 10^{-4}$, 
the value of $g_{b\bar{b}}$ was determined:

\begin{equation}
g_{b\bar{b}}=\frac{f_{d} - ( 1 - g_{c\bar{c}} )\epsilon_{Q} - g_{c\bar{c}}
        \epsilon_{C}}{\epsilon_{B} - \epsilon_{Q}}.
\end{equation}

We obtained
\begin{equation}
g_{b\bar{b}}=(2.44 \pm 0.59) \times 10^{-3},
\end{equation}
where the error is statistical only.

\begin{table}
\centerline{
\begin{tabular}{cc} 
  \hline\hline
  Source & Efficiency ($\%$) \\  
  \hline
  B & $5.28   \pm 0.09$ ($\epsilon_B$) \\
  C & $0.165  \pm 0.018$ ($\epsilon_C$) \\
  Q & $0.00967 \pm 0.00038$ ($\epsilon_Q$) \\
  \hline
\end{tabular}
}
\caption{
\label{Table:SELEFF}
 Selection efficiencies after all cuts for the three categories.
 Errors are statistical only.
}
\end{table}
\vskip 1truecm

\section{Systematic Errors \label{SEC:SYS}}

\vskip .5truecm

\noindent
The efficiencies for the three event categories were evaluated using the 
Monte-Carlo simulation.
The limitations of the simulation in estimating these efficiencies lead to 
an uncertainty on the result.
The error due to limited Monte-Carlo statistics in the efficiency
evaluation was $\Delta g_{b\bar{b}} = \pm 0.12 \times 10^{-3}$.

A large fraction of events remaining after the selection cuts
contain $B$ and $D$ hadrons.
The uncertainty in the knowledge of the physical processes in the simulation
of heavy-flavor production and decays constitutes a source of systematic
error.
All the physical simulation parameters were varied within their allowed 
experimental ranges.
In particular, the $B$ and $D$ hadron lifetimes, their  production rates,
and the mean $B$ hadron energy
were varied following the recommendations of the LEP Heavy
Flavour Working Group \cite{LEPHF}.
We also varied the assumed form of the $B$ energy 
distribution, taking the difference between the optimised Bowler and Peterson
functions~\cite{sldbfrag} to assign an error.
The uncertainties, which are typically small, are summarized in Table \ref{SYSERR}.

\begin{table}
\centerline{
\begin{tabular}{lc} 
  \hline\hline
  Source & $\Delta g_{b\bar{b}}$ $(10^{-3})$ \\
  \hline
  Monte Carlo statistics                        & $\pm 0.12$ \\
  $B$ hadron lifetimes                          & $\pm 0.01$ \\
  $B$ hadron production                         & $\pm 0.02$ \\
  Mean $B$ hadron energy                        & $\pm 0.10$ \\
  $B$ fragmentation function                    & $\pm 0.09$ \\
  $B$ hadron charged multiplicities             & $\pm 0.03$ \\
  $B \to DDX$ fraction                          & $\pm 0.07$ \\
  $D$ hadron lifetimes                          & $\pm 0.01$ \\
  $D$ hadron production                         & $\pm 0.02$ \\
  $D$ hadron charged multiplicities             & $\pm 0.02$ \\
  Theoretical modelling of $g$ \ra \bb kinematics       & $\pm 0.24$ \\
  $b$ quark mass                                & $\pm 0.05$ \\
  $g_{c\bar{c}}$                                & $\pm 0.12$ \\
  Tracking efficiency                           & $\pm 0.04$ \\
  \hline
  Total                                         & $\pm 0.34$ \\
  \hline
\end{tabular}
}
\caption{
\label{SYSERR}
  Systematic uncertainties on $g_{b\bar{b}}$.
}
\end{table}

\begin{figure}[h]      
\centerline{\epsfxsize 2.5 truein \epsfbox{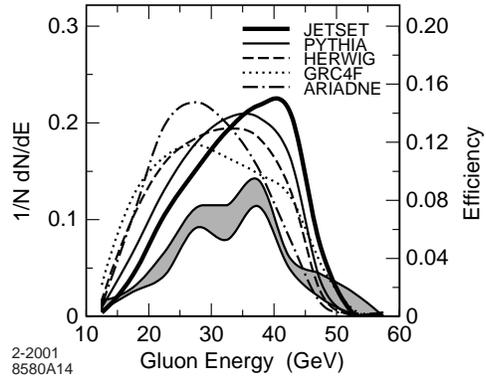}}
\caption[]{Energy distribution (left-hand scale) of gluons for the $g$ \ra \bb process in
different models (see text). Our calculated efficiency and its uncertainty (see text)
are shown as the band (right-hand scale).
}
\label{fig:eg}
\end{figure}

The dominant systematic error arises via the
signal tagging efficiency, $\epsilon_B$, from the dependence on 
modelling the kinematics of the split gluon:
its energy $E_g$, its mass $m_g$ and the decay angle, $\theta^*$, of
the two $B$ hadrons in their center-of-mass frame relative to the 
gluon direction.
In our default Monte Carlo simulation the kinematics of the 
signal events are based on the JETSET parton shower,
 which is in good agreement with the theoretical predictions
\cite{seymour}.
In order to estimate the uncertainty on this assumption, we have
investigated alternative models, namely
the PYTHIA~6.136, HERWIG~6.1~\cite{Marchesini:1992ch}, and ARIADNE 4.08~\cite{ariadne}
models, as well as the exact leading-order matrix-element calculation 
GRC4F~\cite{Fujimoto:1997wj}. 
In each case, we generated, at the parton level, events containing $g$ \ra \bb. 
For illustration the $E_g$ distributions are shown in Fig.~\ref{fig:eg}. 
The efficiency function computed with JETSET was then reweighted by 
the ratio of the new model to JETSET initial distributions to obtain 
a new estimate of the average efficiency. The resulting $g_{b\bar{b}}$
values are shown in Table~\ref{Table:newgbb}.
We took the central value of this ensemble, $g_{b\bar{b}}$ = 2.44 $\times10^{-3}$, 
as our central result, and
assigned a systematic error of $\Delta g_{b\bar{b}} = \pm 0.24 \times 10^{-3}$
based on the full range of values. 
For illustration the efficiency and its uncertainty are
shown in Fig.~\ref{fig:eg}. 

\begin{table}
\centerline{
\begin{tabular}{cc}
  \hline\hline
  Model &  $g_{b\bar{b}}$ ($\times10^{-3}$) \\   \hline
JETSET 7.4	&  2.20 \\
PYTHIA 6.136	&  2.22 \\
HERWIG 6.1	&  2.50 \\
ARIADNE 4.08    &  2.41 \\ 
GRC4F     	&  2.68 \\ \hline  
\end{tabular}
}
\caption{$g_{b\bar{b}}$ values resulting from different models of the
gluon kinematics.
}
\label{Table:newgbb}
\end{table}

The dependence of the efficiency on the $b$-quark mass was also  
investigated at the generator level using the GRC4F Monte Carlo program.
The variation of $\epsilon_B$ was computed 
for $b$-quark masses between $4.7$ and $5.3$ GeV/c$^2$.
The resultant uncertainty is estimated to be 
$\Delta g_{b\bar{b}} = \pm 0.05 \times 10^{-3}$.
The measured uncertainty in the production fraction of $g$ \ra \cc background events,
$\Delta g_{c\bar{c}} = \pm 0.40{\rm \%}$, gives an error 
$\Delta g_{b\bar{b}} = \pm 0.12 \times 10^{-3}$.

In the Monte-Carlo simulation charged tracks used 
in the topological vertex tag were rejected to reproduce better the 
distributions in the data.
Uncertainties in the efficiencies due to this rejection
were assessed by evaluating the Monte-Carlo efficiencies without the 
 rejection algorithm.
The difference in the $g_{b\bar{b}}$ result was taken as a symmetric 
systematic error, $\Delta g_{b\bar{b}} = \pm 0.04 \times 10^{-3}$.

As cross checks we varied independently the value of $y_{cut}$ used in the 4-jet event
selection and the value of the neural network output, $Y$, used to select
the final event sample. In each case we found results consistent with
those determined using the optimised value.
Table \ref{SYSERR} summarizes the different sources of systematic error
on $g_{b\bar{b}}$.
The total systematic error was estimated to be the sum in quadrature,
$0.34 \times 10^{-3}$.

\vskip 1truecm

\section{Summary} 

\vskip .5truecm
 
\noindent
Using the excellent flavor-tagging capabilities of the SLD tracking system,
and a new technique incorporating a multivariate neural network analysis,
we have measured the probability for gluon splitting into \bb in hadronic
$Z^0$ decays.
The result is
$$
g_{b\bar{b}}=(2.44\pm0.59{\rm (stat.)}\pm0.34{\rm (syst.)})\times10^{-3} 
$$
where the first error is statistical and the second is the sum in quadrature of
systematic effects. 
This represents the most precise determination of $g_{b\bar{b}}$ to date.
Our result is consistent with previous measurements~\cite{GBBLEP}.
It is also consistent with the theoretical expectation
$g_{b\bar{b}}$ = $1.75\pm0.40\times10^{-3}$~\cite{seymour}, where the
central value corresponds to \alpmzsq = 0.118 and a
$b$-quark mass of 5.0 GeV/$c^2$. Finally, we found that the predictions of 
the models PYTHIA6.136 ($g_{b\bar{b}}$ = $1.5\times10^{-3}$) and 
HERWIG6.1 ($g_{b\bar{b}}$ = $2.5\times10^{-3}$),
with their default parameter settings, are consistent with our measurement.
The prediction of ARIADNE4.08 ($g_{b\bar{b}}$ = $1.3\times10^{-3}$) is slightly
below our measurement. 

\section*{Acknowledgments}

We thank Mike Seymour for helpful conversations.
We thank the personnel of the SLAC accelerator department and the technical
staffs of our collaborating institutions for their outstanding efforts
on our behalf. 
This work was supported by the U.S. Department of Energy and National Science
Foundation, the UK Particle Physics and Astronomy Research Council, the
Istituto Nazionale di Fisica Nucleare of Italy, the Japan-US Cooperative
Research Project on High Energy Physics, and the Korea Research Foundation.
 
\vskip 1truecm

\noindent
{\bf $^{**}$ List of Authors}
%
%
%
\begin{center}
\def\iAOMORI{$^{(1)}$}
\def\iBRI{$^{(2)}$}
\def\iBRUN{$^{(3)}$}
\def\iBU{$^{(4)}$}
\def\iCOLO{$^{(5)}$}
\def\iCSU{$^{(6)}$}
\def\iFERR{$^{(7)}$}
\def\iFRAS{$^{(8)}$}
\def\iJHU{$^{(9)}$}
\def\iLBL{$^{(10)}$}
\def\iMASS{$^{(11)}$}
\def\iMISSI{$^{(12)}$}
\def\iMIT{$^{(13)}$}
\def\iMOSCOW{$^{(14)}$}
\def\iNAGO{$^{(15)}$}
\def\iOREG{$^{(16)}$}
\def\iOXF{$^{(17)}$}
\def\iPERU{$^{(18)}$}
\def\iRAL{$^{(19)}$}
\def\iRUTG{$^{(20)}$}
\def\iSLAC{$^{(21)}$}
\def\iSOONG{$^{(22)}$}
\def\iTENN{$^{(23)}$}
\def\iTOHO{$^{(24)}$}
\def\iUCSB{$^{(25)}$}
\def\iUCSC{$^{(26)}$}
\def\iVAND{$^{(27)}$}
\def\iWASH{$^{(28)}$}
\def\iWISC{$^{(29)}$}
\def\iYALE{$^{(30)}$}

  \baselineskip=.75\baselineskip
\mbox{Koya Abe\unskip,\iTOHO}
\mbox{Kenji Abe\unskip,\iNAGO}
\mbox{T. Abe\unskip,\iSLAC}
\mbox{I. Adam\unskip,\iSLAC}
\mbox{H. Akimoto\unskip,\iSLAC}
\mbox{D. Aston\unskip,\iSLAC}
\mbox{K.G. Baird\unskip,\iMASS}
\mbox{C. Baltay\unskip,\iYALE}
\mbox{H.R. Band\unskip,\iWISC}
\mbox{T.L. Barklow\unskip,\iSLAC}
\mbox{J.M. Bauer\unskip,\iMISSI}
\mbox{G. Bellodi\unskip,\iOXF}
\mbox{R. Berger\unskip,\iSLAC}
\mbox{G. Blaylock\unskip,\iMASS}
\mbox{J.R. Bogart\unskip,\iSLAC}
\mbox{G.R. Bower\unskip,\iSLAC}
\mbox{J.E. Brau\unskip,\iOREG}
\mbox{M. Breidenbach\unskip,\iSLAC}
\mbox{W.M. Bugg\unskip,\iTENN}
\mbox{D. Burke\unskip,\iSLAC}
\mbox{T.H. Burnett\unskip,\iWASH}
\mbox{P.N. Burrows\unskip,\iOXF}
\mbox{A. Calcaterra\unskip,\iFRAS}
\mbox{R. Cassell\unskip,\iSLAC}
\mbox{A. Chou\unskip,\iSLAC}
\mbox{H.O. Cohn\unskip,\iTENN}
\mbox{J.A. Coller\unskip,\iBU}
\mbox{M.R. Convery\unskip,\iSLAC}
\mbox{V. Cook\unskip,\iWASH}
\mbox{R.F. Cowan\unskip,\iMIT}
\mbox{G. Crawford\unskip,\iSLAC}
\mbox{C.J.S. Damerell\unskip,\iRAL}
\mbox{M. Daoudi\unskip,\iSLAC}
\mbox{N. de Groot\unskip,\iBRI}
\mbox{R. de Sangro\unskip,\iFRAS}
\mbox{D.N. Dong\unskip,\iMIT}
\mbox{M. Doser\unskip,\iSLAC}
\mbox{R. Dubois\unskip,\iSLAC}
\mbox{I. Erofeeva\unskip,\iMOSCOW}
\mbox{V. Eschenburg\unskip,\iMISSI}
\mbox{E. Etzion\unskip,\iWISC}
\mbox{S. Fahey\unskip,\iCOLO}
\mbox{D. Falciai\unskip,\iFRAS}
\mbox{J.P. Fernandez\unskip,\iUCSC}
\mbox{K. Flood\unskip,\iMASS}
\mbox{R. Frey\unskip,\iOREG}
\mbox{E.L. Hart\unskip,\iTENN}
\mbox{K. Hasuko\unskip,\iTOHO}
\mbox{S.S. Hertzbach\unskip,\iMASS}
\mbox{M.E. Huffer\unskip,\iSLAC}
\mbox{X. Huynh\unskip,\iSLAC}
\mbox{M. Iwasaki\unskip,\iOREG}
\mbox{D.J. Jackson\unskip,\iRAL}
\mbox{P. Jacques\unskip,\iRUTG}
\mbox{J.A. Jaros\unskip,\iSLAC}
\mbox{Z.Y. Jiang\unskip,\iSLAC}
\mbox{A.S. Johnson\unskip,\iSLAC}
\mbox{J.R. Johnson\unskip,\iWISC}
\mbox{R. Kajikawa\unskip,\iNAGO}
\mbox{M. Kalelkar\unskip,\iRUTG}
\mbox{H.J. Kang\unskip,\iRUTG}
\mbox{R.R. Kofler\unskip,\iMASS}
\mbox{R.S. Kroeger\unskip,\iMISSI}
\mbox{M. Langston\unskip,\iOREG}
\mbox{D.W.G. Leith\unskip,\iSLAC}
\mbox{V. Lia\unskip,\iMIT}
\mbox{C. Lin\unskip,\iMASS}
\mbox{G. Mancinelli\unskip,\iRUTG}
\mbox{S. Manly\unskip,\iYALE}
\mbox{G. Mantovani\unskip,\iPERU}
\mbox{T.W. Markiewicz\unskip,\iSLAC}
\mbox{T. Maruyama\unskip,\iSLAC}
\mbox{A.K. McKemey\unskip,\iBRUN}
\mbox{R. Messner\unskip,\iSLAC}
\mbox{K.C. Moffeit\unskip,\iSLAC}
\mbox{T.B. Moore\unskip,\iYALE}
\mbox{M. Morii\unskip,\iSLAC}
\mbox{D. Muller\unskip,\iSLAC}
\mbox{V. Murzin\unskip,\iMOSCOW}
\mbox{S. Narita\unskip,\iTOHO}
\mbox{U. Nauenberg\unskip,\iCOLO}
\mbox{H. Neal\unskip,\iYALE}
\mbox{G. Nesom\unskip,\iOXF}
\mbox{N. Oishi\unskip,\iNAGO}
\mbox{D. Onoprienko\unskip,\iTENN}
\mbox{L.S. Osborne\unskip,\iMIT}
\mbox{R.S. Panvini\unskip,\iVAND}
\mbox{C.H. Park\unskip,\iSOONG}
\mbox{I. Peruzzi\unskip,\iFRAS}
\mbox{M. Piccolo\unskip,\iFRAS}
\mbox{L. Piemontese\unskip,\iFERR}
\mbox{R.J. Plano\unskip,\iRUTG}
\mbox{R. Prepost\unskip,\iWISC}
\mbox{C.Y. Prescott\unskip,\iSLAC}
\mbox{B.N. Ratcliff\unskip,\iSLAC}
\mbox{J. Reidy\unskip,\iMISSI}
\mbox{P.L. Reinertsen\unskip,\iUCSC}
\mbox{L.S. Rochester\unskip,\iSLAC}
\mbox{P.C. Rowson\unskip,\iSLAC}
\mbox{J.J. Russell\unskip,\iSLAC}
\mbox{O.H. Saxton\unskip,\iSLAC}
\mbox{T. Schalk\unskip,\iUCSC}
\mbox{B.A. Schumm\unskip,\iUCSC}
\mbox{J. Schwiening\unskip,\iSLAC}
\mbox{V.V. Serbo\unskip,\iSLAC}
\mbox{G. Shapiro\unskip,\iLBL}
\mbox{N.B. Sinev\unskip,\iOREG}
\mbox{J.A. Snyder\unskip,\iYALE}
\mbox{H. Staengle\unskip,\iCSU}
\mbox{A. Stahl\unskip,\iSLAC}
\mbox{P. Stamer\unskip,\iRUTG}
\mbox{H. Steiner\unskip,\iLBL}
\mbox{D. Su\unskip,\iSLAC}
\mbox{F. Suekane\unskip,\iTOHO}
\mbox{A. Sugiyama\unskip,\iNAGO}
\mbox{S. Suzuki\unskip,\iNAGO}
\mbox{M. Swartz\unskip,\iJHU}
\mbox{F.E. Taylor\unskip,\iMIT}
\mbox{J. Thom\unskip,\iSLAC}
\mbox{E. Torrence\unskip,\iMIT}
\mbox{T. Usher\unskip,\iSLAC}
\mbox{J. Va'vra\unskip,\iSLAC}
\mbox{R. Verdier\unskip,\iMIT}
\mbox{D.L. Wagner\unskip,\iCOLO}
\mbox{A.P. Waite\unskip,\iSLAC}
\mbox{S. Walston\unskip,\iOREG}
\mbox{A.W. Weidemann\unskip,\iTENN}
\mbox{E.R. Weiss\unskip,\iWASH}
\mbox{J.S. Whitaker\unskip,\iBU}
\mbox{S.H. Williams\unskip,\iSLAC}
\mbox{S. Willocq\unskip,\iMASS}
\mbox{R.J. Wilson\unskip,\iCSU}
\mbox{W.J. Wisniewski\unskip,\iSLAC}
\mbox{J.L. Wittlin\unskip,\iMASS}
\mbox{M. Woods\unskip,\iSLAC}
\mbox{T.R. Wright\unskip,\iWISC}
\mbox{R.K. Yamamoto\unskip,\iMIT}
\mbox{J. Yashima\unskip,\iTOHO}
\mbox{S.J. Yellin\unskip,\iUCSB}
\mbox{C.C. Young\unskip,\iSLAC}
\mbox{H. Yuta\unskip.\iAOMORI}

\it
  \vskip \baselineskip                   
  \baselineskip=.75\baselineskip   
\iAOMORI
  Aomori University, Aomori, 030 Japan, \break
\iBRI
  University of Bristol, Bristol, United Kingdom, \break
\iBRUN
  Brunel University, Uxbridge, Middlesex, UB8 3PH United Kingdom, \break
\iBU
  Boston University, Boston, Massachusetts 02215, \break
\iCOLO
  University of Colorado, Boulder, Colorado 80309, \break
\iCSU
  Colorado State University, Ft. Collins, Colorado 80523, \break
\iFERR
  INFN Sezione di Ferrara and Universita di Ferrara, I-44100 Ferrara, Italy,
\break
\iFRAS
  INFN Laboratori Nazionali di Frascati, I-00044 Frascati, Italy, \break
\iJHU
  Johns Hopkins University,  Baltimore, Maryland 21218-2686, \break
\iLBL
  Lawrence Berkeley Laboratory, University of California, Berkeley, California
94720, \break
\iMASS
  University of Massachusetts, Amherst, Massachusetts 01003, \break
\iMISSI
  University of Mississippi, University, Mississippi 38677, \break
\iMIT
  Massachusetts Institute of Technology, Cambridge, Massachusetts 02139, \break
\iMOSCOW
  Institute of Nuclear Physics, Moscow State University, 119899 Moscow, Russia,
\break
\iNAGO
  Nagoya University, Chikusa-ku, Nagoya, 464 Japan, \break
\iOREG
  University of Oregon, Eugene, Oregon 97403, \break
\iOXF
  Oxford University, Oxford, OX1 3RH, United Kingdom, \break
\iPERU
  INFN Sezione di Perugia and Universita di Perugia, I-06100 Perugia, Italy,
\break
\iRAL
  Rutherford Appleton Laboratory, Chilton, Didcot, Oxon OX11 0QX United Kingdom,
\break
\iRUTG
  Rutgers University, Piscataway, New Jersey 08855, \break
\iSLAC
  Stanford Linear Accelerator Center, Stanford University, Stanford, California
94309, \break
\iSOONG
  Soongsil University, Seoul, Korea 156-743, \break
\iTENN
  University of Tennessee, Knoxville, Tennessee 37996, \break
\iTOHO
  Tohoku University, Sendai, 980 Japan, \break
\iUCSB
  University of California at Santa Barbara, Santa Barbara, California 93106,
\break
\iUCSC
  University of California at Santa Cruz, Santa Cruz, California 95064, \break
\iVAND
  Vanderbilt University, Nashville,Tennessee 37235, \break
\iWASH
  University of Washington, Seattle, Washington 98105, \break
\iWISC
  University of Wisconsin, Madison,Wisconsin 53706, \break
\iYALE
  Yale University, New Haven, Connecticut 06511. \break

\rm
%

\end{center}

\vfill\eject 

\begin{thebibliography}{99}
\bibitem{quark}
In this analysis we do not distinguish between quark and antiquark jets.
\bibitem{alpha}
The peak at 1 in the distribution of $|\cos\alpha_{1234}|$ is an artefact of
the mapping from the distribution of $\alpha_{1234}$.
\end{thebibliography}
\end{document}